# Fine-UNETR for PSMA PET/CT Lesion Segmentation: Automated Tumor Quantification and Overall Survival Stratification in Prostate Cancer

**Mansour Abtahi, PhD[1], Chae Moon Hong, MD, PhD[1, 2], Nikhil Deveshwar, PhD[1], Stellamaris Nwihim, MPH[1], Minkyung Lee, MD, PhD[1], Peder E.Z. Larson, PhD[1], and Thomas A. Hope, MD[1,*]**

[1] Department of Radiology and Biomedical Imaging, University of California San Francisco, San Francisco, CA, 94158

[2] Department of Nuclear Medicine, Kyungpook National University, Republic of Korea, 41944

Corresponding author: Thomas Hope, E-mail: Thomas.Hope@ucsf.edu, Phone: 415-221-4810 ext. 22648, 185 Berry Street Bldg B, #395, San Francisco CA 94107

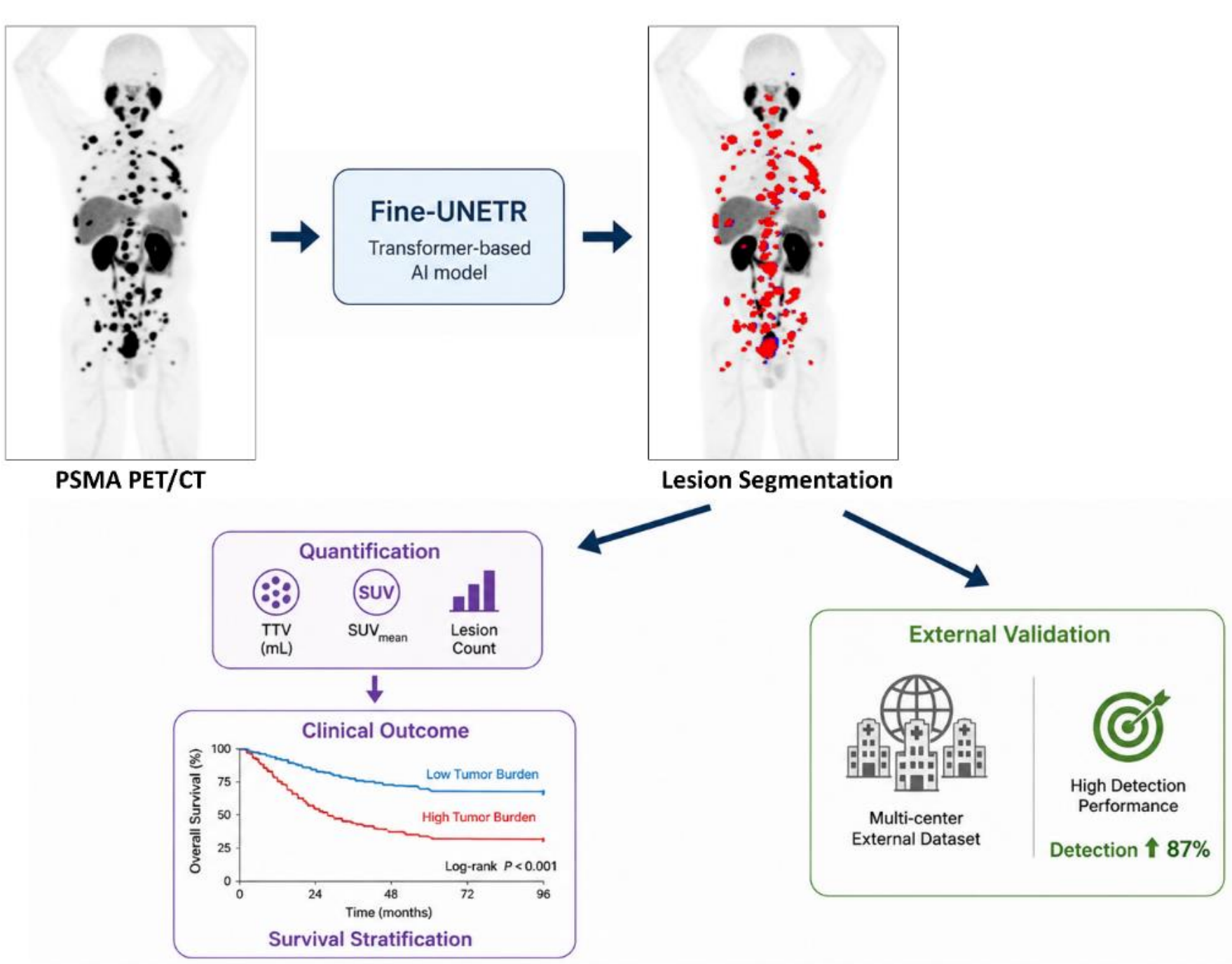

Graphical Abstract

**Abstract:**

**Background/Introduction:** To develop and evaluate Fine-UNETR, a Vision Transformer-based architecture for automated segmentation of PSMA-avid lesions on whole-body PET/CT, and to assess the clinical utility of AI-derived tumor burden biomarkers for overall survival stratification in patients undergoing radioligand therapy.

**Methods:** In this retrospective study, 373 PSMA PET/CT scans (mean age, 71±8 years) from patients with prostate cancer were analyzed. Fine-UNETR, a modified UNETR with 8×8×8 voxel patch embedding and axial sliding window training, was trained on 299 scans and validated on 74 scans. Segmentation performance was evaluated using Dice similarity coefficient, sensitivity, precision, and lesion detection rate. AI-derived biomarkers were compared with expert ground truth using Pearson correlation. Overall survival stratification was assessed in an independent cohort of 67 pre-radioligand therapy patients using Kaplan-Meier analysis and log-rank testing. External validation was performed on 192 cases from the AutoPET IV PSMA PET/CT dataset.

**Results:** Fine-UNETR achieved a Dice similarity coefficient (DSC) of 66.63%, sensitivity of 70.27%, precision of 67.77%, and a lesion detection rate of 79.53% (96.05% for lesions with SUVmax ≥ 5). On the external validation dataset, the model achieved a DSC of 44.11% and a lesion detection rate of 87.18%, indicating that lesion detection performance was preserved despite reduced voxel-level overlap. AI-derived biomarkers showed excellent agreement with ground truth (total tumor volume: r=0.984; total lesion uptake: r=0.989; lesion count: r=0.960). In the clinical cohort, total tumor volume (p=0.0019), SUVmax (p=0.014), and SUVmean (p=0.016) significantly stratified overall survival.

**Conclusion:** Fine-UNETR enables accurate automated whole-body PSMA lesion segmentation and reliable tumor burden quantification. Performance on an external multi-center dataset demonstrates robustness in lesion detection rate despite evidence of domain shift. AI-derived biomarkers significantly stratified overall survival in a pre-radioligand therapy cohort, supporting the clinical utility of automated PSMA PET/CT quantification for prognostication.

## 1 Introduction

Prostate cancer is the second most commonly diagnosed cancer in men worldwide and the most frequently diagnosed cancer in men in the United States, accounting for 29% of all new male cancer cases, and remains a leading cause of cancer-related death [1]. While localized diseases are often curable with surgery or radiation, a significant proportion of patients develop metastatic disease, which remains incurable and is managed with systemic therapies including androgen deprivation, chemotherapy, and more recently, radioligand therapy [2,3]. Accurate assessment of metastatic burden is critical for treatment selection, response monitoring, and prognostication [4].

Prostate-specific membrane antigen (PSMA) is a type II transmembrane glycoprotein that is highly overexpressed on prostate cancer cells, particularly in metastatic and castration-resistant disease, making it an ideal target for molecular imaging and therapy [5]. PSMA-targeted positron emission tomography/computed tomography (PET/CT) using radiotracers such as [$^{68}$Ga]Ga-PSMA-11 and [$^{18}$F]-labeled PSMA agents has revolutionized the staging and restaging of prostate cancer, offering superior sensitivity and specificity compared to conventional imaging modalities [6–8]. Furthermore, PSMA-derived quantitative biomarkers such as total tumor volume (TTV), total lesion uptake, and lesion count have emerged as prognostic indicators and predictors of response to PSMA-targeted radioligand therapy with $^{177}$Lu-PSMA [9,10]. However, manual segmentation of PSMA-avid lesions across whole-body scans is time-consuming, subject to inter-observer variability, and impractical for routine clinical workflows or large-scale quantitative biomarker extraction [11,12].

Early approaches to automated PSMA PET/CT quantification employed semiautomatic methods combining SUV-based thresholding with anatomical masks to exclude physiological uptake in organs such as kidneys, liver, and salivary glands [11,13,14]. While these techniques demonstrated feasibility for whole-body tumor burden assessment, they required manual corrections and were limited by fixed threshold values that may not generalize across heterogeneous lesion populations. The advent of deep learning has enabled substantial progress in automated medical image segmentation. Convolutional neural networks (CNNs), particularly U-Net architectures and their 3D variants, have been increasingly applied to PSMA PET/CT lesion segmentation. Initial work by Zhao et al. demonstrated the potential of deep neural networks using a 2.5D U-Net architecture for lesion characterization in the pelvic region [15]. Subsequent studies expanded to whole-body segmentation using the self-configuring nnU-Net framework [16,17]. Various architectural innovations have been explored, including weighted batch-wise dice loss functions to improve detection of small metastases [18], attention mechanisms for uptake classification [19], and maximum intensity projection-based approaches that leverage 2D representations to enhance 3D segmentation accuracy [20]. More recent work has investigated adaptive voxel-weighted loss functions using L1 norms to address class imbalance challenges inherent in metastatic lesion segmentation [21]. These studies have reported Dice similarity coefficients (DSC) ranging from 0.44 to 0.70 and lesion detection rates of 73–93% depending on SUVmax thresholds, with performance generally improving for lesions with SUVmax ≥ 5. However, CNN-based approaches face inherent limitations: they possess limited receptive fields that may fail to capture the spatial relationships between anatomically distant lesions, such as the co-occurrence patterns of skeletal metastases across the axial skeleton, and most implementations rely on computationally intensive sliding window inference at full resolution.

Despite these advances, the detection and segmentation of small metastatic lesions remain a persistent challenge. Small lesions occupy only a few voxels after standard preprocessing and resampling, making them susceptible to being missed or under-segmented, particularly when they exhibit lower uptake closer to background. More recently, Vision Transformer (ViT) architectures have been explored for PSMA PET/CT analysis. Yazdani et al. employed Swin UNETR for the segmentation of both lesions and organs at risk using 100 labeled scans, demonstrating improved segmentation compared to CNN-based methods [22]. ViTs have emerged as a powerful alternative to traditional CNN architectures, demonstrating superior performance in natural image recognition tasks through self-attention mechanisms that model global context [23]. The UNETR (U-Net Transformers) architecture combines ViT encoders with CNN decoders, leveraging the complementary strengths of both paradigms [24]. While ViT-based models have shown promise in medical imaging applications including organ segmentation and tumor detection [25,26], their application to PSMA PET/CT lesion segmentation remains largely unexplored.

In this study, we present Fine-UNETR, a modified UNETR architecture with fine-grained 8×8×8 voxel patch embedding optimized for detecting small and low uptake metastatic lesions in whole-body PSMA PET/CT imaging. By reducing the patch size, Fine-UNETR preserves finer spatial detail critical for detecting small metastases while maintaining the global context modeling capabilities of the transformer architecture. To address the computational challenges of processing high-resolution whole-body scans while maintaining spatial coherence, we introduce an axial sliding window training strategy that extracts overlapping patches along the superior-inferior axis while preserving full transverse spatial context. We hypothesize that the fine-grained transformer architecture combined with our axial sliding window approach will achieve improved whole-body lesion segmentation.

## 2 Materials and Methods

### *2.1 Patient Cohort, Image Preprocessing, and Clinical Evaluation*

This retrospective study was approved by the institutional review board at our institution, and informed consent was waived due to the retrospective nature of the analysis. Two independent cohorts, both obtained between 2016 and 2024, were analyzed (Table 1): a segmentation cohort of 373 selected PSMA PET/CT scans for model development (299 training, 74 validation), and a clinical evaluation cohort of 67 pre-radioligand therapy (pre-RLT) PSMA PET/CT scans not included in model training or validation.

All PET and CT volumes were coregistered using rigid registration and resampled to 4.0×4.0×3.27mm spacing. PET activity concentrations were converted to standardized uptake values (SUV) and clipped to [0, 50], while CT images were windowed to [−1000, 1200] Hounsfield Units; both were linearly scaled to [0, 1].

**Table 1.** Patient Cohort and Imaging Characteristics

| Characteristic | Value | Characteristic | Value |
|---|---|---|---|
| **Segmentation Cohort:** | | **Clinical Evaluation Cohort:** | |
| Number of scans | 373 | Number of scans | 67 |
| Training / Validation split | 299 / 74 | Age, mean ± SD | 72.4 ± 8.9 |
| Age, mean ± SD | 71 ± 8 | Age range | 54–93 |
| Age range | 44–93 | Sex | male |
| Sex | male | | |
| **PET Acquisition:** | | **CT Acquisition:** | |
| Reconstruction method | 3D iterative (OSEM-based) | Tube voltage, kVp | 100–140 (120 kVp in 88%) |
| Matrix size | 128×128 to 440×440 | Tube current modulation | Automatic |
| Slice thickness, mm (median) | 2–5 (3.27) | Matrix size | 512×512 |
| **Radiotracer:** | | **Scanner Manufacturer:** | |
| $^{68}$Ga-PSMA-11 | 76% | GE Healthcare | 49% |
| $^{18}$F-labeled PSMA | 24% | Siemens Healthineers | 27% |
| Injected activity, mean ± SD, MBq | 238 ± 83 | Philips Healthcare | 22% |
| Uptake time, mean ± SD, min | 63 ± 9 | United Imaging Healthcare | 2% |
| **Most Common Scanner Models:** | | | |
| GE Discovery STE | 42% | | |
| Philips Vereos PET/CT | 17% | | |
| Siemens Biograph64 Vision 600 | 12% | | |

Ground-truth annotations were manually delineated by a nuclear medicine physician (15 years of experience) using MIM software (MiM Software, USA). Top part of Figure 1 presents one sample with fused PET/CT coronal and sagittal slices, the corresponding PET maximum intensity projection (MIP) and ground truth masks. An initial SUV threshold of ≥ 3.0 was applied; for liver metastases, a threshold of 1.5 times the mean SUV of normal liver was used. The SUV cutoff was adjusted downward in 0.5 decrements when suspected lesions were not adequately captured. Unlike studies using a fixed SUV ≥ 3 threshold [16,27,28], we relied on expert delineation based on relative tracer uptake to capture clinically relevant lesions below fixed thresholds. To address class imbalances, cases with small-volume metastases (<1 mL total lesion volume), hepatic metastases, and head/neck lesions were oversampled four times in the training set. The validation set consisted of 74 unique cases, without augmentation, to ensure unbiased performance evaluation.

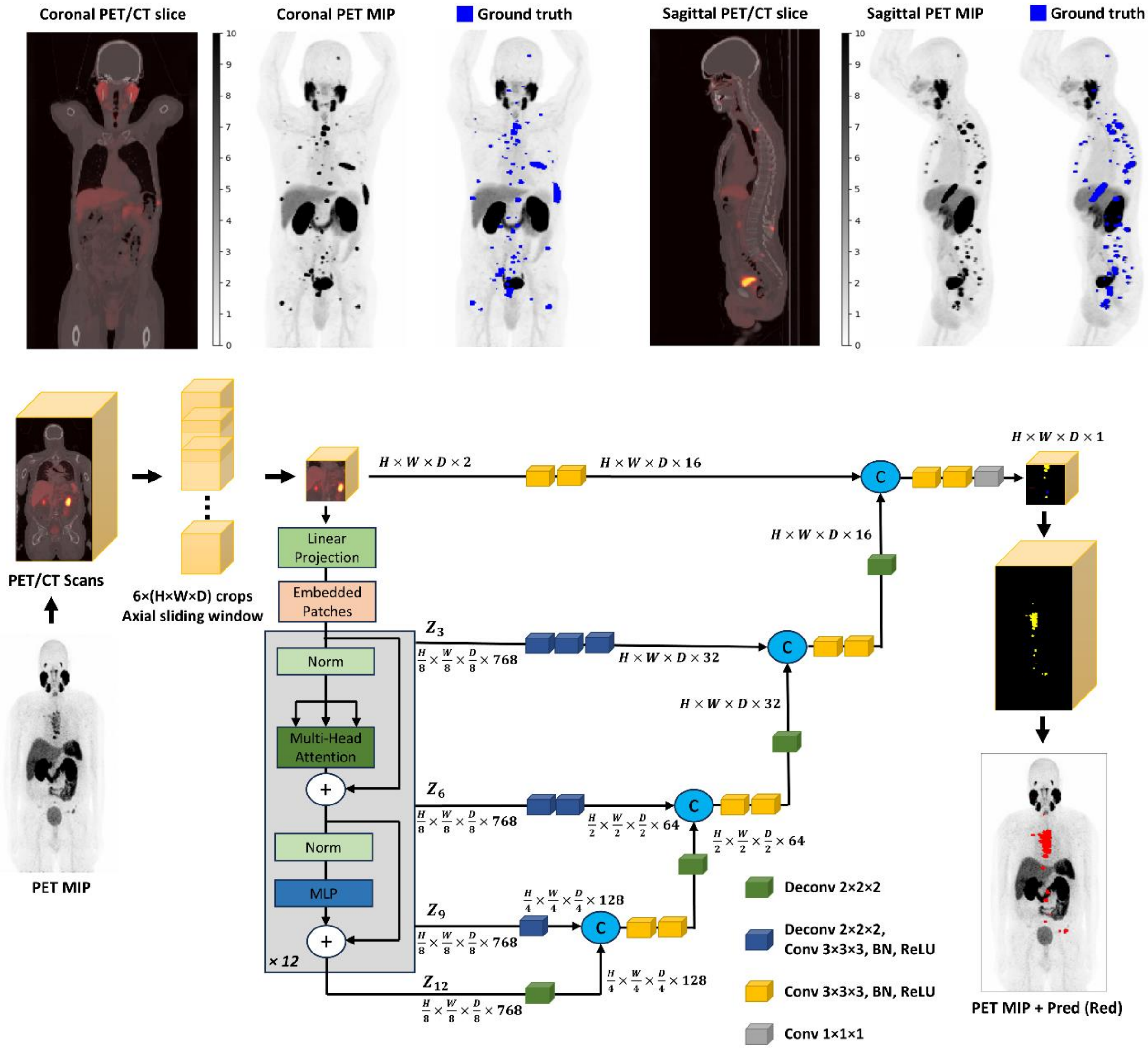

Figure 1. Representative PSMA PET/CT case and Fine-UNETR architecture overview. *Top:* Coronal (left) and sagittal (right) views of a patient with extensive metastatic prostate cancer. PET maximum intensity projection (MIP) displaying whole-body radiotracer distribution, and ground truth lesion annotations (blue) superimposed on the PET MIP. This case demonstrates widespread osseous metastatic disease involving the spine, pelvis, and ribs, illustrating the complexity of whole-body lesion segmentation. *Bottom:* Overview of the Fine-UNETR architecture. The network processes dual-channel PET and CT volumes using an axial sliding window strategy that generates 6 overlapping crops per scan. Input patches are divided into fine-grained 8×8×8 voxel tokens. The Vision Transformer (ViT) encoder consists of 12 sequential blocks. Feature maps are extracted at four depths ($Z_3$, $Z_6$, $Z_9$, $Z_{12}$) and passed to the convolutional decoder via skip connections. Skip connection features are concatenated with decoder features at each resolution level. MIP of a PET scan (bottom left) and corresponding model predictions overlaid in red (bottom right).

For clinical evaluation, overall survival (OS) was analyzed using Kaplan-Meier curves in a separate, unseen dataset of 67 pre-RLT patients using AI-derived segmentations and associated PET parameters. All AI-derived segmentations used for clinical evaluation were generated fully automatically without any physician review, correction, or manual editing. Patients were stratified by median values of whole-body SUVmax and TTV, with a separate analysis comparing patients with SUVmean ≥ 10 against the remainder. Follow-up duration was calculated from the date of pre-RLT PET/CT to final follow-up. All statistical analyses were performed using R, and differences between the Kaplan-Meier curves were assessed using the log-rank test. A p-value of $< 0.05$ was considered statistically significant.

*2.2 External Validation Dataset*

External validation was performed using the publicly available AutoPET IV dataset [29], which consists of multi-center whole-body PET/CT scans acquired across multiple institutions and scanner platforms. This dataset includes heterogeneous imaging protocols and patient populations, providing a robust benchmark for assessing model generalizability. The data set also includes multiple PSMA tracers ([$^{18}$F]-labeled as well as [$^{68}$Ga]Ga-PSMA-11), and these are known to have different biodistributions. Since our training did not include all [$^{18}$F]-labeled tracers found in AutoPET and primarily consisted of [$^{68}$Ga]Ga-PSMA-11 scans, only AutoPET cases imaged with [$^{68}$Ga]Ga-PSMA-11 were included in the external validation analysis.

Additional filtering steps were applied to improve the consistency of external validation. Cases with suspected annotation inconsistencies were excluded following visual inspection, as variations in labeling quality could bias performance assessment. Furthermore, the head and neck regions of the ground truth labels were excluded from analysis because these areas were removed or truncated in the PET/CT dataset to preserve patient anonymity, resulting in incomplete ground truth annotations. After applying these criteria, a total of 192 cases were included for external validation. All images were preprocessed using the same pipeline as the training dataset, including resampling, intensity normalization, and spatial alignment, and inference was performed without any model retraining or fine-tuning. These preprocessing steps were applied to ensure fair evaluation and avoid confounding factors related to tracer differences and incomplete annotations.

### *2.3 Deep Learning Architecture*

We implemented a modified UNETR [24] architecture integrating a Vision Transformer (ViT) encoder with a convolutional decoder (Figure 1, bottom). Fine-UNETR uses a fine-grained 8×8×8 voxel patch size, smaller than the standard 16×16×16, yielding 1,728 patches per 96×96×96 input crop compared to 216 in conventional UNETR. The network takes dual-channel PET and CT volumes as input for end-to-end 3D lesion segmentation. Each patch is projected into a 768-dimensional embedding space with learnable positional embeddings and processed through 12 transformer blocks, each comprising multi-head self-attention (12 heads) and a feed-forward network (hidden dimension 3,072) with instance normalization and residual connections. This design enables global contextual modeling across whole-body PET/CT volumes. The convolutional decoder reconstructs segmentation maps using multi-scale features from transformer layers 3, 6, 9, and 12 via skip connections, progressively upsampled through four stages to produce a voxel-wise lesion probability map.

Overall, the Fine-UNETR architecture contains approximately 110 million trainable parameters, with the majority (about 85 million) located in the Vision Transformer encoder. This fine-grained transformer-based design enables accurate modeling of subtle PSMA uptake patterns while preserving global anatomical context across the entire imaging volume. We compare the proposed Fine-UNETR directly with the standard UNETR architecture using the same training setup, input resolution, and evaluation metrics to assess the impact of the finer 8×8×8 patch size, which increases the number of tokens from 216 to 1,728, on segmentation performance.

### *2.4 Axial Sliding Window Training Strategy*

To enable whole-body PET/CT processing within GPU memory constraints, we employed an axial sliding window (ASW) strategy (Figure 1, bottom). For each volume, six partially overlapping 96×96×96 voxel crops are extracted along the superior–inferior (Z) axis, sharing the same transverse X–Y center selected using a lesion-aware sampling strategy that includes at least one segmented lesion voxel in the union of six crops to increase exposure to PSMA-avid regions. The starting positions along the Z-axis are distributed to span the full craniocaudal extent of the body with controlled overlap between adjacent crops.

All six crops are processed simultaneously as a batch. During training and validation, crop-level predictions are merged along the Z-axis using uniform averaging in overlapping regions, producing a continuous prediction map. The loss is computed on the reconstructed axial volume rather than individual crops, providing consistent supervision across the entire superior–inferior extent and enabling the model to learn relationships between body regions. This strategy reduces memory usage approximately eightfold compared with full volume processing while preserving full transverse spatial context.

*2.5 Training Configuration and Evaluation Metrics*

We employed two loss functions: global Dice loss (Equation 1), which directly optimizes segmentation overlap using squared predictions to penalize overconfident errors, and the Instance-Center-aware Iterative (ICI) loss [30] (Equation 2), which combines global Dice ($a = 0.25$), instance-wise Dice ($b = 0.5$), and center-wise Dice ($c = 0.25$) to improve detection of multiple small lesions. Instance-wise Dice evaluates per-lesion quality via connected component matching, while center-wise Dice focuses supervision on 7×7×7 voxel lesion centers.

$$D_{global} = 1 - \frac{2\sum_i p_i g_i + \varepsilon}{\sum_i p_i^2 + \sum_i g_i^2 + \varepsilon} \tag{1}$$

$$L_{ICI} = 1 - a \times D_{global} - b \times D_{instance} - c \times D_{center} \tag{2}$$

Where $p_i$ represents predicted probabilities, $g_i$ ground truth binary labels, and $\varepsilon = 1\times10^{-5}$ is a smoothing term to prevent division by zero.

Training used AdamW (weight decay 0.05) with linear warmup (50 epochs to $1\times10^{-4}$) followed by cosine annealing to $5\times10^{-5}$ over 350 epochs. Models were trained on two NVIDIA A6000 GPUs (48

GB each) using distributed data parallel with batch size 1 volume per GPU (6 crops per volume) and 16-bit mixed precision. Training required approximately 110 and 195 GPU-hours for 400 epochs using Dice and ICI loss, respectively.

Model performance was assessed at both voxel and lesion levels. Image-wise Dice similarity coefficient (DSC), sensitivity, and precision were computed per scan. Lesion-wise evaluation used connected component analysis to match predicted and ground truth lesions based on spatial overlap, with DSC, sensitivity, and precision averaged across matched pairs. The lesion detection rate was defined as the percentage of ground truth lesions with at least one overlapping prediction. All metrics used a probability threshold of 0.5 for binarization.

## 3 Results

**Table 2.** Performance of different experiments

| Metrics | Fine-UNETR ASW Dice Loss | Fine-UNETR ASW ICI Loss | Fine-UNETR Random crops Dice Loss | UNETR ASW Dice Loss |
|---|---|---|---|---|
| Mean image-wise DSC | 66.63 | 66.11 | 64.17 | 65.62 |
| Mean image-wise sensitivity | 70.27 | 68.89 | 69.07 | 66.50 |
| Mean image-wise precision | 67.77 | 69.32 | 65.29 | 68.98 |
| Mean lesion detection rate | 79.53 | 73.59 | 75.21 | 76.11 |
| Mean lesion-wise DSC | 46.59 | 46.50 | 45.71 | 46.86 |
| Mean lesion-wise sensitivity | 56.72 | 51.87 | 52.26 | 53.43 |
| Mean lesion-wise precision | 48.50 | 49.25 | 48.31 | 49.58 |

ASW, Axial sliding window; DSC, Dice Similarity Coefficient; ICI, Instance-Center-aware Iterative.

Quantitative results are summarized in Table 2. Fine-UNETR with ASW strategy and Dice loss achieved the highest image-wise DSC (66.63%), sensitivity (70.27%), and lesion detection rate (79.53%) among all configurations. Dice loss outperformed ICI loss for detection rate (79.53% vs. 73.59%), while ICI loss yielded slightly higher precision. The ASW strategy improved DSC (66.63% vs. 64.17%) and detection rate (79.53% vs. 75.21%) over random cropping. Fine-UNETR also outperformed standard UNETR in sensitivity (70.27% vs. 66.50%) and detection rate (79.53% vs. 76.11%).

Table 3 presents performance stratified by lesion SUVmax. Model performance improved consistently with increasing SUVmax thresholds, likely due to the enhanced conspicuity of PSMA-avid lesions. Detection rate increased from 79.53% for all lesions to 96.05% for SUVmax ≥ 5, with lesion-wise DSC improving from 46.59% to 63.04%. These results demonstrate that the model reliably detects lesions with high radiotracer uptake.

**Table 3.** Performance of the Fine-UNETR model on lesions with varying SUVmax values.

| Metrics | All lesions | Lesions with SUVmax ≥ 2 | Lesions with SUVmax ≥ 3 | Lesions with SUVmax ≥ 5 |
|---|---|---|---|---|
| Mean image-wise DSC | 66.63 | 66.72 | 66.97 | 68.61 |
| Mean image-wise sensitivity | 70.27 | 70.43 | 70.95 | 72.26 |
| Mean image-wise precision | 67.77 | 67.77 | 68.01 | 69.61 |
| Mean lesion detection rate | 79.53 | 84.53 | 91.98 | 96.05 |
| Mean lesion-wise DSC | 46.59 | 48.22 | 53.84 | 63.04 |
| Mean lesion-wise sensitivity | 56.72 | 58.54 | 63.59 | 72.61 |
| Mean lesion-wise precision | 48.50 | 50.65 | 56.66 | 63.82 |

DSC, Dice Similarity Coefficient.

Figure 2 illustrates segmentation results for four representative cases with varying disease burden. False negatives (orange arrows) primarily involved low-SUVmax lesions, consistent with Table 3. Figure 3 shows strong correlations between predicted and ground truth biomarkers: TTV ($r = 0.984$), total lesion uptake (calculated as the sum of SUV values across all segmented lesion voxels; $r = 0.989$), lesion count (defined as the number of connected components in the binary segmentation mask; $r = 0.960$), and SUVmean ($r = 0.894$). Figure 4 shows that DSC correlated positively with tumor volume ($r = 0.528$, $p < 0.001$), while detection rate showed no significant correlation ($r = -0.173$, $p = 0.141$), indicating consistent lesion identification regardless of disease extent.

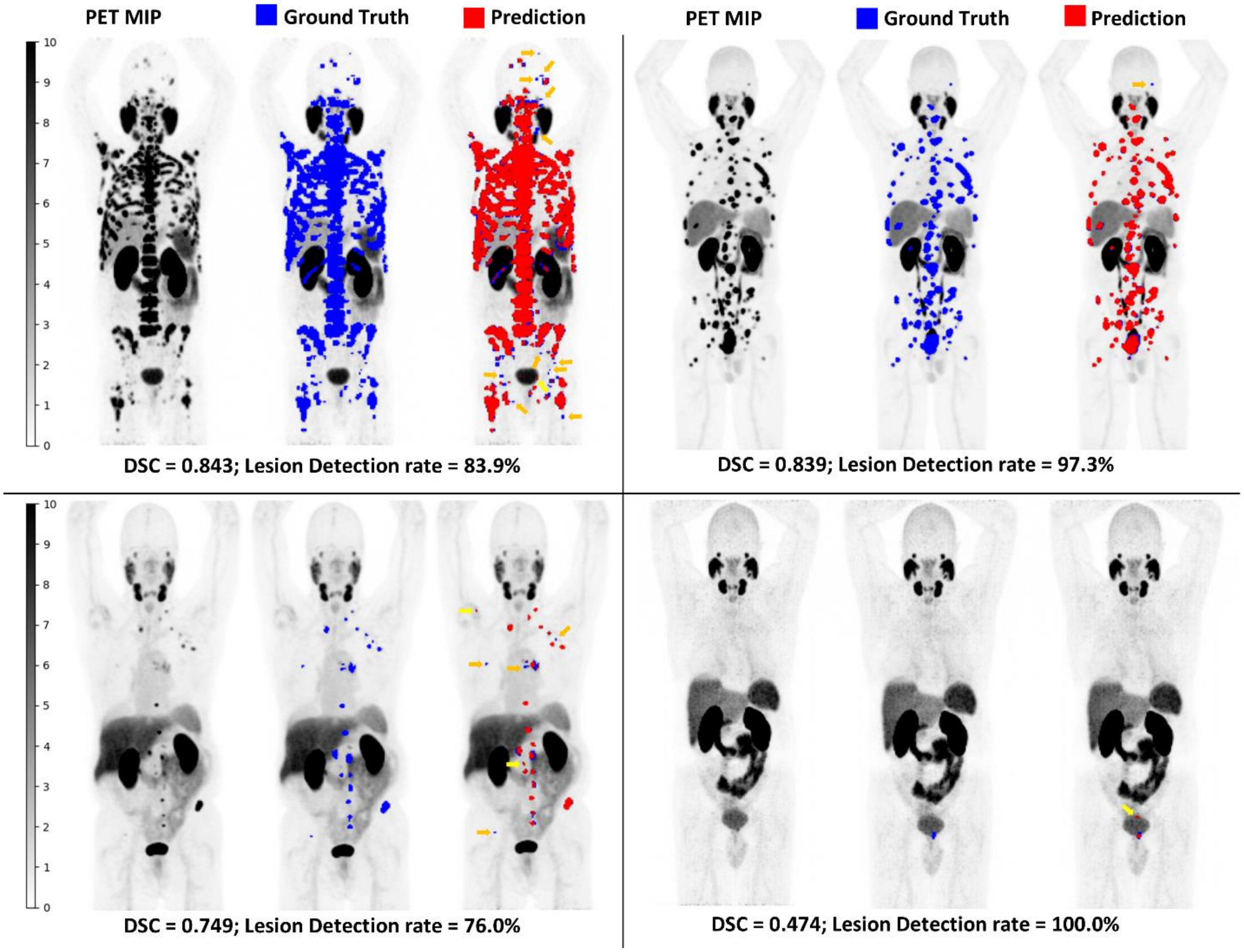

Figure 2. Visual comparison of lesion segmentation across four representative cases. Each panel displays (left to right): PET maximum intensity projection (MIP), expert ground truth annotations (blue), and model predictions (red). The color bar indicates standardized uptake values (SUV, range 0-10). Performance metrics are shown below each case: Dice similarity coefficient (DSC) and lesion-wise detection rate. Orange arrows indicate false negative lesions, while yellow arrows highlight false positive predictions. Top row: Cases with extensive poly-metastatic disease involving the skeleton, demonstrating high segmentation performance (DSC=0.843, 0.839) and detection rates (83.9%, 97.3%). Bottom left: Case with moderate disease burden showing good agreement between predictions and ground truth (DSC=0.749, detection rate=76.0%). Bottom right: Case represents an oligometastatic pattern with only one lesion achieving 100% lesion detection despite lower DSC (0.474) because of the false positive predictions.

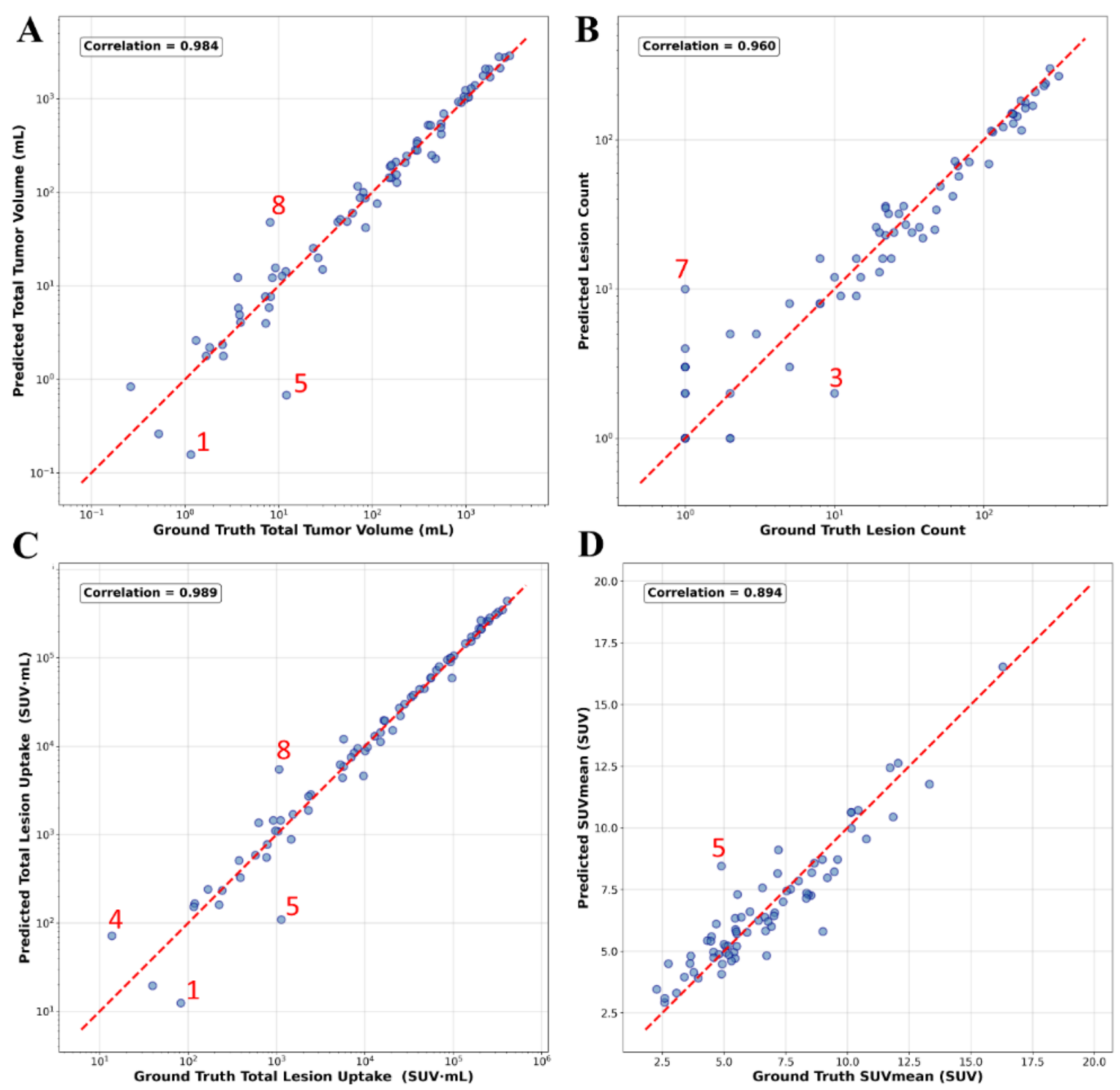

Figure 3. Correlation analysis between predicted and ground truth quantitative PSMA PET biomarkers. Scatter plots comparing Fine-UNETR automated segmentation-derived metrics with manual annotations for (A) total tumor volume (mL) showing excellent agreement (r=0.984), (B) lesion count demonstrating strong correlation (r=0.960), (C) total lesion uptake (TLU, SUV·mL) with the highest correlation (r=0.989), and (D) whole-body mean standardized uptake value (SUVmean) showing good correlation (r=0.894). The red numbers indicate specific cases discussed in the text.

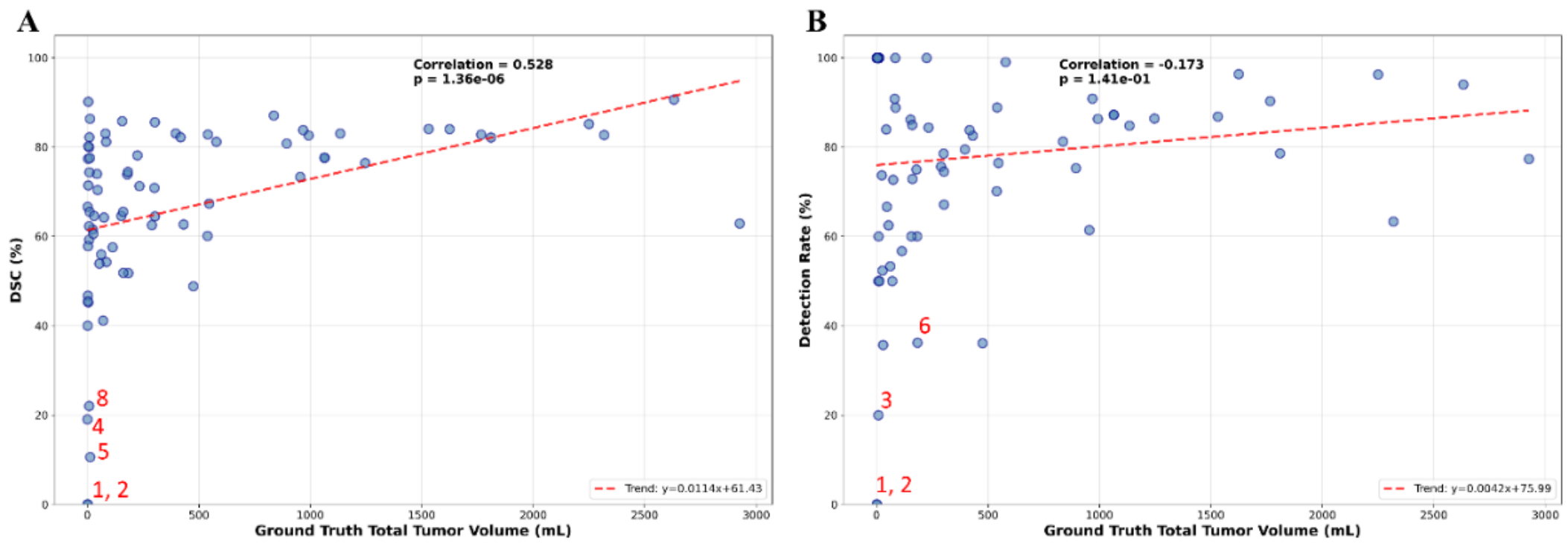

Figure 4. Relationship between segmentation performance and disease burden. (A) Image-wise Dice similarity coefficient (DSC) and (B) lesion detection rate as a function of ground truth total tumor volume. Red dashed lines indicate linear regression trend lines with equations shown. Correlation coefficients (r) and p-values are displayed for each plot. (A) DSC shows a moderate positive correlation with tumor volume (r=0.528, p<0.001), indicating that segmentation overlap metrics improve with increasing disease burden. (B) Detection rate shows no significant correlation with tumor volume (r=-0.173, p=0.141), demonstrating that the model maintains consistent lesion detection performance regardless of disease extent. Red numbered points indicate cases with notable performance deviations discussed in the text.

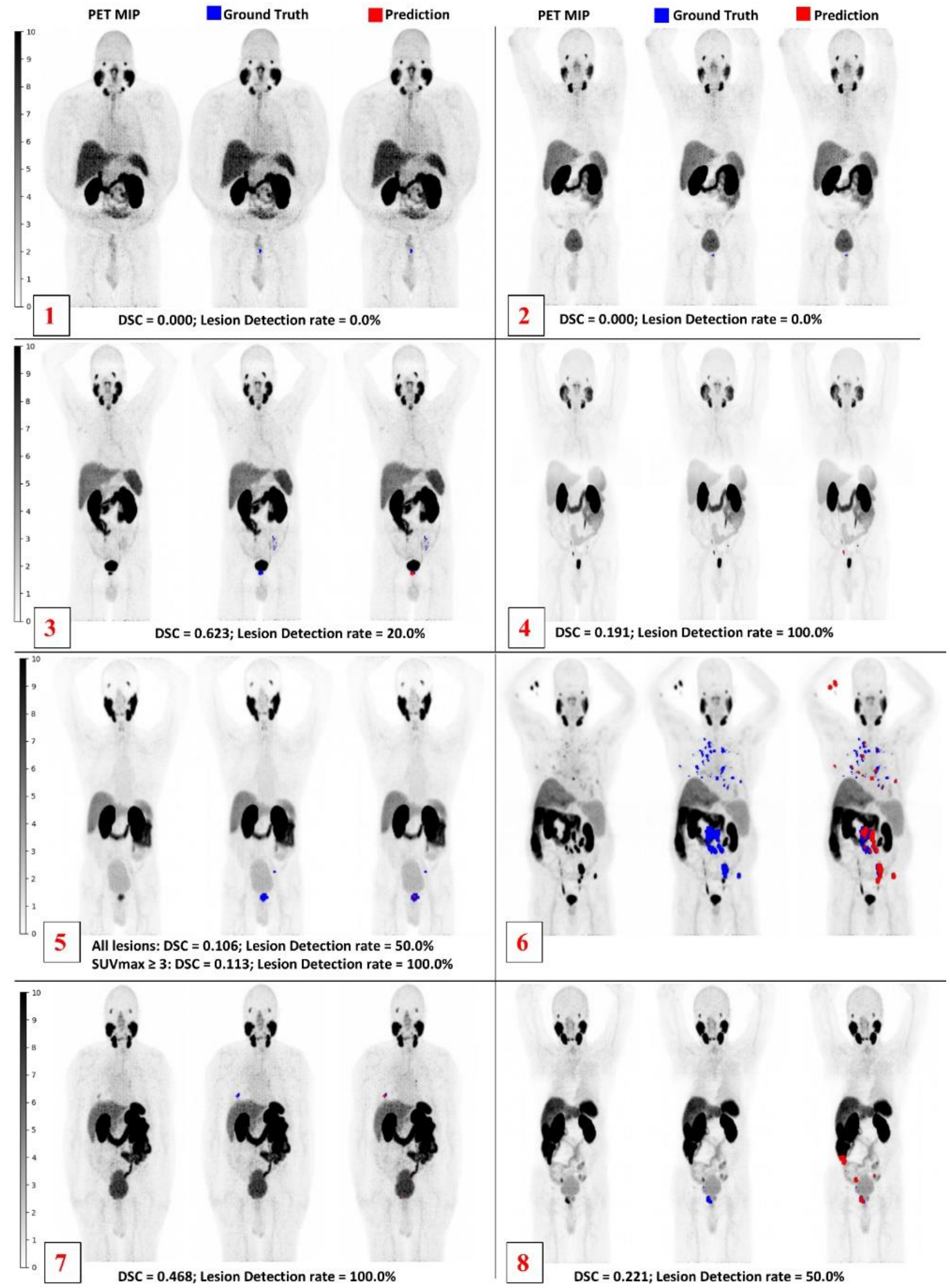

**Figure 5.** Analysis of challenging cases with suboptimal segmentation performance. Eight outlier cases identified from quantitative analysis in Figures 3 and 4 (numbered annotations in red). Each panel displays (left to right): PET maximum intensity projection (MIP), expert ground truth annotations (blue), and model predictions (red). The color bar indicates standardized uptake values (SUV, range 0–10). Cases 1 and 2: Complete segmentation failures (DSC = 0.000, detection rate = 0.0%). Both cases contain small, low-intensity lesions with subtle PSMA uptake that are difficult to distinguish from background noise, representing borderline findings that may also challenge human interpretation. Case 3: Partial detection (DSC = 0.623, detection rate = 20.0%) where the model successfully segmented the most conspicuous lesion but missed the majority of smaller foci with low PSMA uptake. Case 4: High detection rate (100.0%) but low DSC (0.191) due to a false positive prediction in the ureter region, a common site of physiological PSMA uptake from radiotracer excretion. Cases 5 and 6: Performance metrics are shown for all lesions and for lesions with SUVmax ≥ 3 separately. Case 5 achieved only 50.0% detection rate for all lesions but 100.0% when restricted to lesions with SUVmax ≥ 3. Similarly, Case 6 improved from 36.2% to 93.8% detection rate when considering only higher-uptake lesions, demonstrating that the model reliably detects lesions with higher PSMA uptake while missing subtle, low-uptake foci. Case 7: Achieved 100.0% detection rate but low DSC (0.468) due to false positive segmentations. Case 8: Shows 50.0% detection rate and low DSC (0.221) with overestimation of total tumor volume from false positive predictions.

To better understand the model performance on challenging cases, Figure 5 presents eight challenging outlier cases identified from Figures 3 and 4. Our analysis of challenging cases revealed two primary limitations. First, some lesions with low SUVmax values remain challenging to detect, as evidenced by the dramatic improvement in detection rate when filtering by SUVmax threshold (79.53% for all lesions vs. 96.05% for SUVmax ≥ 5). This limitation mirrors the challenges faced by human readers when interpreting subtle uptake in the context of background noise and physiological activity, and has been consistently reported across prior studies (18,20). Notably, many low-intensity lesions may represent findings of uncertain clinical significance, and the model's tendency to miss these cases may align with clinical practice, where such findings often require correlation with other imaging or follow-up rather than immediate intervention. Second, false positive predictions, particularly in regions of physiological PSMA uptake such as the ureters from radiotracer excretion or near the injection site, can affect quantitative metrics.

External validation results using the AutoPET dataset are summarized in Table 4. The Fine-UNETR model achieved a mean image-wise DSC of 44.11% and a lesion detection rate of 87.18% across all lesions when applied to this dataset. Performance improved for lesions with higher uptake, with detection rates rising from 87.18% for all lesions to 96.15% for lesions with SUVmax ≥ 5, accompanied by increases in lesion-wise DSC (34.76% to 43.41%) and sensitivity (53.67% to 67.00%).

**Table 4.** Performance of the Fine-UNETR model on lesions with different SUVmax on the external validation dataset.

| Metrics | All lesions | Lesions with SUVmax ≥ 2 | Lesions with SUVmax ≥ 3 | Lesions with SUVmax ≥ 5 |
|---|---|---|---|---|
| Mean image-wise DSC | 44.11 | 44.11 | 44.72 | 47.25 |
| Mean image-wise sensitivity | 42.33 | 42.33 | 60.73 | 67.47 |
| Mean image-wise precision | 59.52 | 59.52 | 52.27 | 50.63 |
| Mean lesion detection rate | 87.18 | 88.41 | 92.40 | 96.15 |
| Mean lesion-wise DSC | 34.76 | 35.00 | 37.06 | 43.41 |
| Mean lesion-wise sensitivity | 53.67 | 54.01 | 56.59 | 67.00 |
| Mean lesion-wise precision | 40.55 | 41.03 | 43.57 | 45.74 |

DSC, Dice Similarity Coefficient.

Compared with the internal validation dataset, the model demonstrated a reduction in voxel-level overlap (DSC: 66.63% vs. 44.11%) but an increase in lesion detection rate (79.53% vs. 87.18%). The decrease in DSC likely reflects **differences** in annotation protocols such as variations in lesion boundary definitions, and the exclusion of anatomically truncated regions (e.g., head and neck) in the external dataset. In contrast, the improved detection rate, particularly for lesions with higher SUVmax, suggests that the model retains strong sensitivity for identifying clinically significant PSMA-avid lesions in external data.

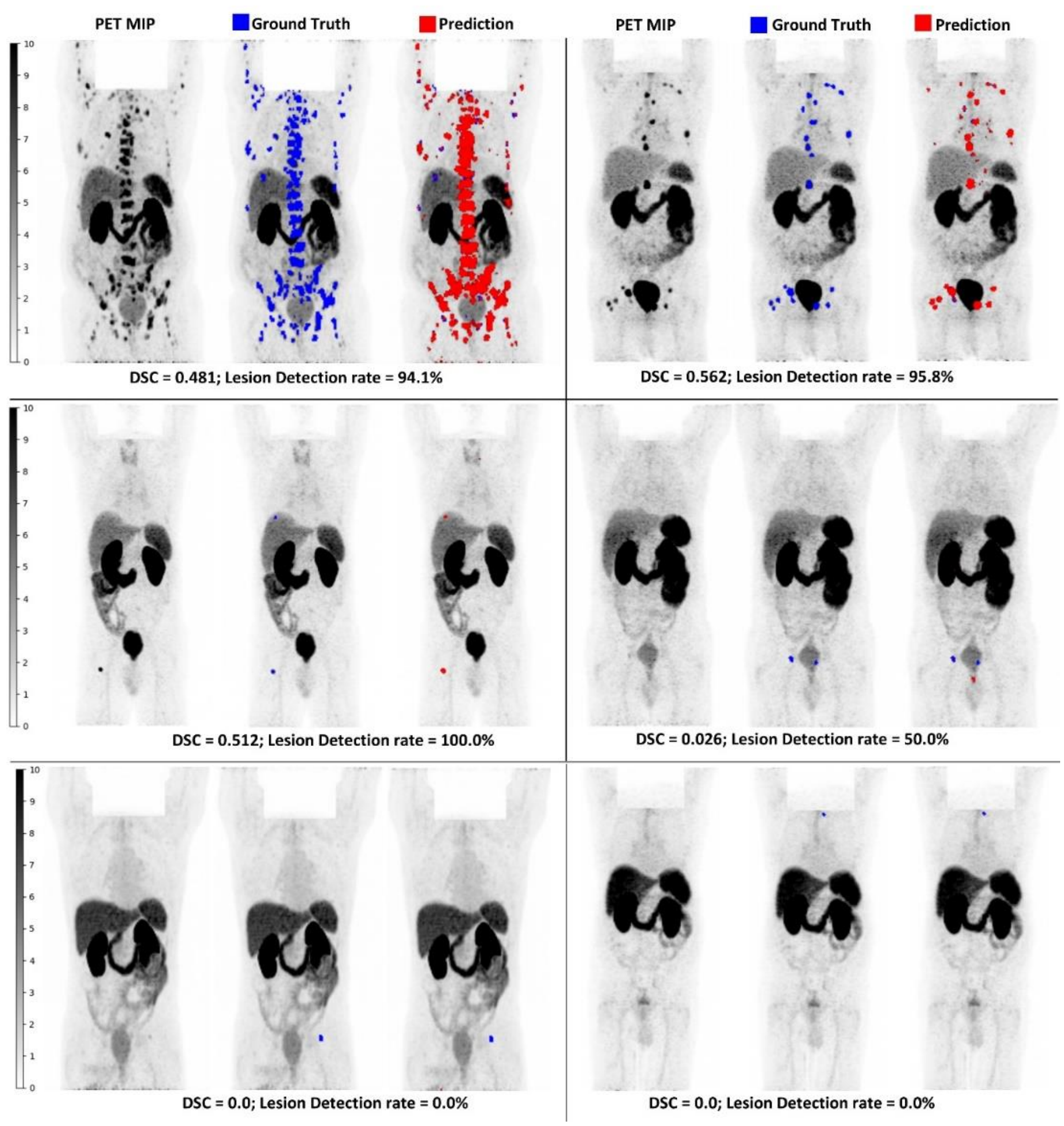

Figure 6. External validation examples demonstrating segmentation performance across representative and challenging cases. Each panel shows (left to right) PET maximum intensity projection (MIP), expert annotations (blue), and model predictions (red). The color bar indicates standardized uptake values (SUV, 0–10), and performance metrics (Dice similarity coefficient [DSC] and lesion detection rate) are shown below each case. Top row: High disease burden cases with strong detection and good spatial agreement (DSC = 0.481–0.562; detection rate = 94.1–95.8%). Middle left: Limited disease burden with accurate localization and 100% detection (DSC = 0.512). Remaining cases: Challenging low-intensity or sparse lesions with reduced performance, including partial detection (DSC = 0.026; detection rate = 50.0%) and complete failures (DSC = 0.000; detection rate = 0.0%).

Figure 6 illustrates representative examples of segmentation performance on the external validation dataset, including both well-performing and challenging cases. In cases with high disease burden and prominent PSMA uptake, the model typically achieved high lesion detection rates with moderate to good spatial agreement. In contrast, performance decreased in cases with low disease burden or subtle uptake patterns, as shown. These qualitative findings are consistent with the quantitative results in Table 4, indicating that model performance is strongly influenced by lesion conspicuity in the external validation dataset.

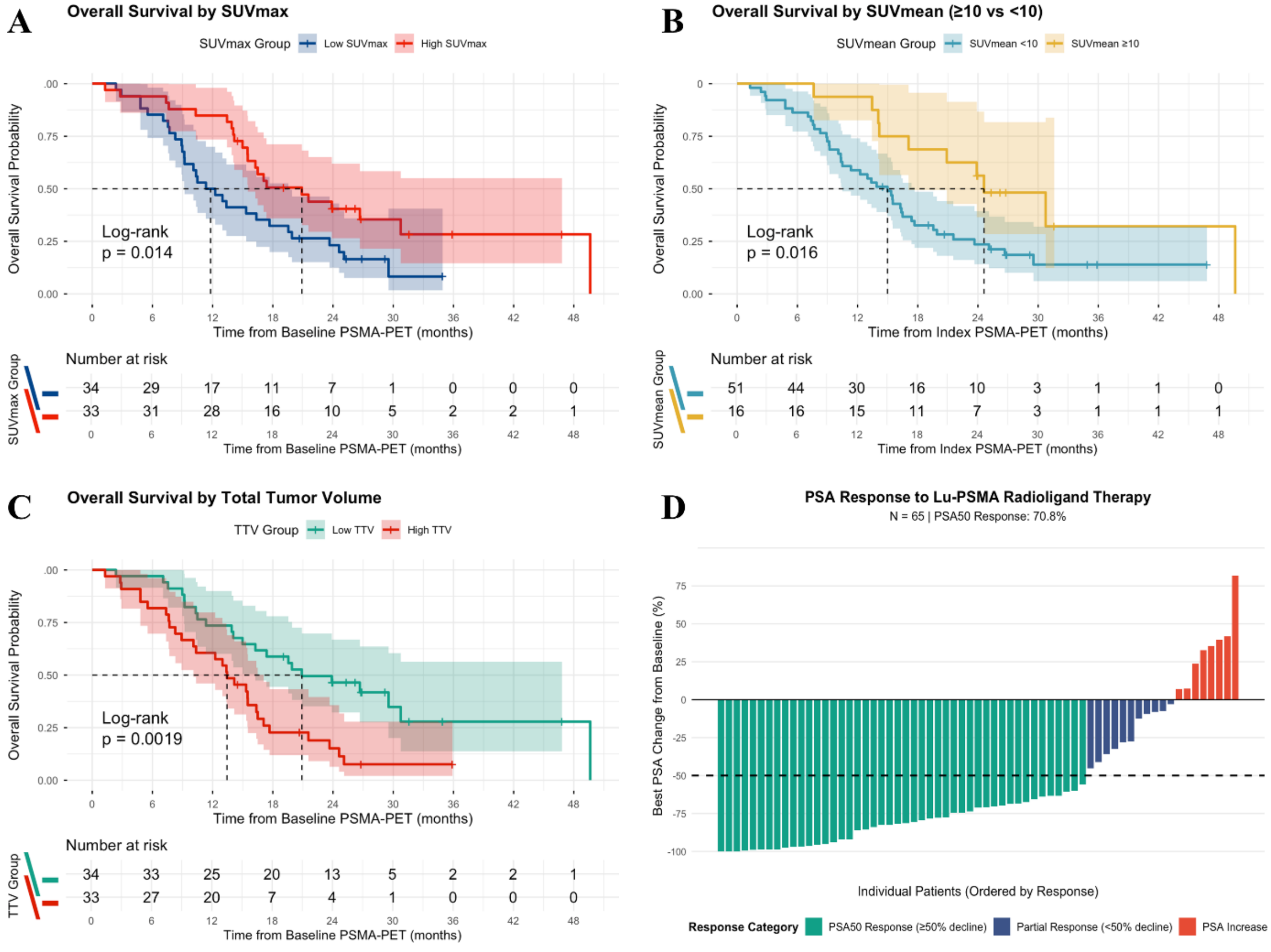

Figure 7. Clinical evaluation of model-derived PET parameters in the independent pre-radioligand therapy cohort (n=67). Kaplan-Meier survival curves stratified by (A) whole-body SUVmax, showing significantly shorter overall survival in the low SUVmax group (median OS: 11.8 vs. 20.9 months; log-rank p=0.014), (B) whole-body SUVmean (≥10 vs. <10), demonstrating poorer survival in the SUVmean <10 group (median OS: 15.0 vs. 24.6 months; log-rank p=0.016), and (C) total tumor volume (TTV), showing significantly worse outcomes in the high TTV group (median OS: 13.4 vs. 20.9 months; log-rank p=0.0019). (D) Waterfall plot of best PSA response to Lu-PSMA radioligand therapy, with a PSA50 response rate of 70.8%.

In the clinical evaluation cohort (n=67), the model-derived PET parameters were significantly associated with OS. The model-derived PET parameters showed that the median [interquartile ranges] for SUVmax, SUVmean, and TTV were 46.99 [29.13–74.62], 7.16 [5.26–9.69], and 514.5 [202.9–1201.5] mL. Poorer OS was significantly linked to low SUVmax (median OS: 11.8 vs. 20.9 months; p=0.014; Figure 7A), SUVmean < 10 (median OS: 15.0 vs. 24.6 months; p=0.016; Figure 7B) and high TTV (median OS: 20.9 vs. 13.4 months; p=0.0019; Figure 7C).

## 4 Discussion

We developed Fine-UNETR for automated PSMA PET/CT lesion segmentation, achieving a DSC of 66.63% and lesion detection rate of 79.53%. The fine-grained 8×8×8 patch embedding and ASW strategy were designed to generalize across patients with low and high TTV, improve small lesion detection and enable whole-body analysis. We further demonstrated that AI-derived parameters are prognostic for overall survival in pre-RLT patients.

These results compare favorably to prior PSMA PET segmentation methods [16–18,22]. The finer patch size improved sensitivity (70.27% vs. 66.50% image-wise; 56.72% vs. 53.43% lesion-wise) and detection rate (79.53% vs. 76.11%) over standard UNETR, likely by better capturing small metastatic foci. The increased computational cost was offset by the ASW strategy, which reduced memory requirements approximately eightfold. Unlike random cropping used in nnU-Net [16,17], ASW processes six overlapping patches spanning the full craniocaudal extent, enabling the network to learn contextual relationships between body regions, from head to pelvis, relevant for distinguishing metastatic patterns. This whole-body awareness is particularly important for PSMA imaging, where lesion distribution across skeletal, nodal, and visceral compartments carries diagnostic and prognostic significance. This proved superior to random cropping for both DSC (66.63% vs. 64.17%) and detection rate (79.53% vs. 75.21%).

Importantly, we did not apply a fixed SUV cutoff for annotation, instead relying on expert delineation based on relative uptake and anatomical correlation [16,27,28]. This captures clinically relevant lesions below standard thresholds, creating a more challenging but comprehensive dataset. For lesions with SUVmax $\geq 3$ and $\geq 5$, detection rates reached 91.98% and 96.05%, respectively, indicating near-complete identification of highly avid metastases most relevant for treatment planning and response assessment. The strong biomarker correlations, particularly for TTV (r=0.984) and total lesion uptake (r=0.989), support clinical utility for disease burden quantification. The slightly lower SUVmean correlation (r=0.894) reflects sensitivity of intensity-based metrics to boundary errors.

Our automated segmentation achieves sufficient accuracy to support these clinical applications, potentially enabling routine quantitative assessment that would be impractical with manual delineation. To validate this clinical utility, we applied the model to an independent cohort (n=67) and demonstrated that model-derived parameters significantly stratified overall survival. Our findings are consistent with the follow-up data from the TheraP trial, which demonstrated that an SUVmean $\geq 10$ is a significant predictor of overall survival [31]. And these findings also align with prior work showing that PSMA-derived tumor volume predicts outcomes after $^{177}$Lu-PSMA radioligand therapy [9,10].

External validation on [$^{68}$Ga]Ga-PSMA-11 data in the AutoPET dataset further supports the generalizability of Fine-UNETR across heterogeneous multi-center data. Despite differences in imaging protocols, scanners, patient populations, and annotation methods, the model maintained strong lesion detection performance (87.18%), exceeding that of the internal validation cohort (79.53%). Voxel-level overlap decreased, likely due to annotation variability, differences in lesion boundary definitions, and anatomical truncation of regions such as the head and neck in the external dataset. Nevertheless, consistently high detection rates, particularly for higher SUVmax lesions, indicate robust identification of clinically relevant PSMA-avid lesions under domain shift. These findings suggest that while boundary delineation is dataset-dependent, lesion localization generalizes well across institutions.

Several limitations should be acknowledged. First, although external validation was performed using a multi-center dataset, the evaluation was restricted to cases using [$^{68}$Ga]Ga-PSMA-11, and extending this work to [$^{18}$F]-labeled PSMA tracers such as PSMA-1007 remains to be established.

Second, ground truth annotations were performed by a single expert reader, and inter-observer variability in lesion delineation, particularly for subtle findings, was not assessed. Third, the model was trained and evaluated on cases with visible disease, and its performance on truly negative scans requires further investigation. Future directions include integration of anatomical segmentation to reduce false positives in organs with physiological uptake and exploration of multi-task learning frameworks that jointly predict lesion segmentation and quantitative biomarkers. The extension of this approach to other PSMA-targeted radiotracers and therapeutic applications, including dosimetry $^{177}$Lu-PSMA dosimetry, represents an important clinical translation opportunity.

## 5 Conclusion

Fine-UNETR with axial sliding window sampling enables automated whole-body PSMA PET/CT lesion segmentation with high lesion detection performance and reliable tumor burden quantification. Performance on an external multi-center dataset demonstrates robust lesion detection despite domain shift. Fine-UNETR provides accurate and generalizable identification of clinically significant PSMA-avid lesions and enables automated extraction of prognostic imaging biomarkers associated with overall survival. This approach has the potential to support reproducible quantitative assessment and facilitate clinical decision-making in metastatic prostate cancer.

**Declaration of competing interests:**

This work was supported by an investigator-initiated study funded by GE Healthcare. The funding source had no role in study design, data collection, analysis, interpretation, or manuscript preparation.

Thomas Hope reports grant funding to his institution from Bayer, GE Healthcare, Lantheus, Janssen, Novartis, Telix Pharmaceuticals, the Prostate Cancer Foundation, and the National Cancer Institute (R01CA235741). He has received personal fees from AstraZeneca, Bayer, Cardinal Health, BlueEarth Diagnostics, Lantheus, Molecular Partners, Novartis, RayzeBio/BMS, and Sanofi, and reports fees from and equity interest in Curium, AdvanCell, and Utter Therapeutics.

Chae Moon Hong reports research grants from Futurechem and Neurophet; consulting fees from GE Healthcare, Neurophet, and IPSEN Korea; and lecture fees from Siemens Healthineers.

**Funding sources:**

This work was funded by an investigator-initiated study funded by GE Healthcare.

**Data statement:**

The datasets are not publicly available because of institutional privacy restrictions. External validation data is available through the AutoPET IV challenge repository.